\documentclass{article}
\usepackage{spconf,amsmath,graphicx}

\usepackage{enumitem}
\setlist{nosep, leftmargin=14pt}
\usepackage{cite}
\usepackage{amssymb,amsfonts}
\usepackage{algorithmic}
\usepackage{textcomp}
\usepackage{xcolor}
\usepackage{hyperref}
\usepackage{makecell}
\usepackage{booktabs}
\usepackage{multirow}

\title{Full-length-body CBCT imaging in upright position with robotic-arm system: a simulation study}

\name{
Tong Lin$^{1,*}$, Tianling Lyu$^{2,*}$, Zhan Wu$^{1}$, Yan Xi$^{3}$, Wentao Zhu$^{2, \dag}$, Yang Chen$^{1,4,5,\dag}$
\thanks{$^{*}$Tong Lin and Tianling Lyu contribute equally to the paper. }
\thanks{$^{\dag}$ Corresponding authors. Email: \url{chenyang.list@seu.edu.cn} (Yang Chen) and \url{wentao.zhu@zhejianglab.com} (Wentao Zhu). }
}
\address{1. Laboratory of Image Science and Technology, Southeast University, Nanjing, China. \\
2. Research Center of Augmented Intelligence, Zhejiang Lab, Hangzhou, China. \\
3. Shanghai First-Imaging Tech, Shanghai, China. \\
4. Jiangsu Provincial Joint International Research Laboratory of Medical \\
Information Processing, Southeast University, Nanjing, China. \\
5. Key Laboratory of New Generation Artificial Intelligence Technology and Its \\
Interdisciplinary Applications, Southeast University, Nanjing, China}

\begin{document}

\maketitle

\begin{abstract}
Upright position CT scans make it possible for full-length-body imaging at conditions more relevant to daily situations, but the substantial weight of the upright CT scanners increases the risks to floor’s stability and patients’ safety. Robotic-arm CBCT systems are supposed to be a better solution for this task, but such systems still face challenges including long scanning time and low reconstruction quality. To address the above challenges, this paper proposes a novel method to calculate optimal scanning pitch based on data completeness analysis, which can complete the whole-body scan in the shortest time without a significant decline in image quality. Besides, an FDK-style reconstruction method based on normalized projections is proposed to obtain fast image reconstruction. Extensive experiments prove the effectiveness of the proposed optimal scanning trajectory. Qualitative and quantitative comparisons with FDK and iterative algorithms show that the proposed reconstruction method can obtain high imaging quality with reasonable computation costs. The method proposed in this paper is expected to promote the application of robotic-arm CBCT systems in orthopedic functional analysis.

\end{abstract}


\section{Introduction}

\begin{figure}[t]
\centerline{\includegraphics[width=0.4\textwidth]{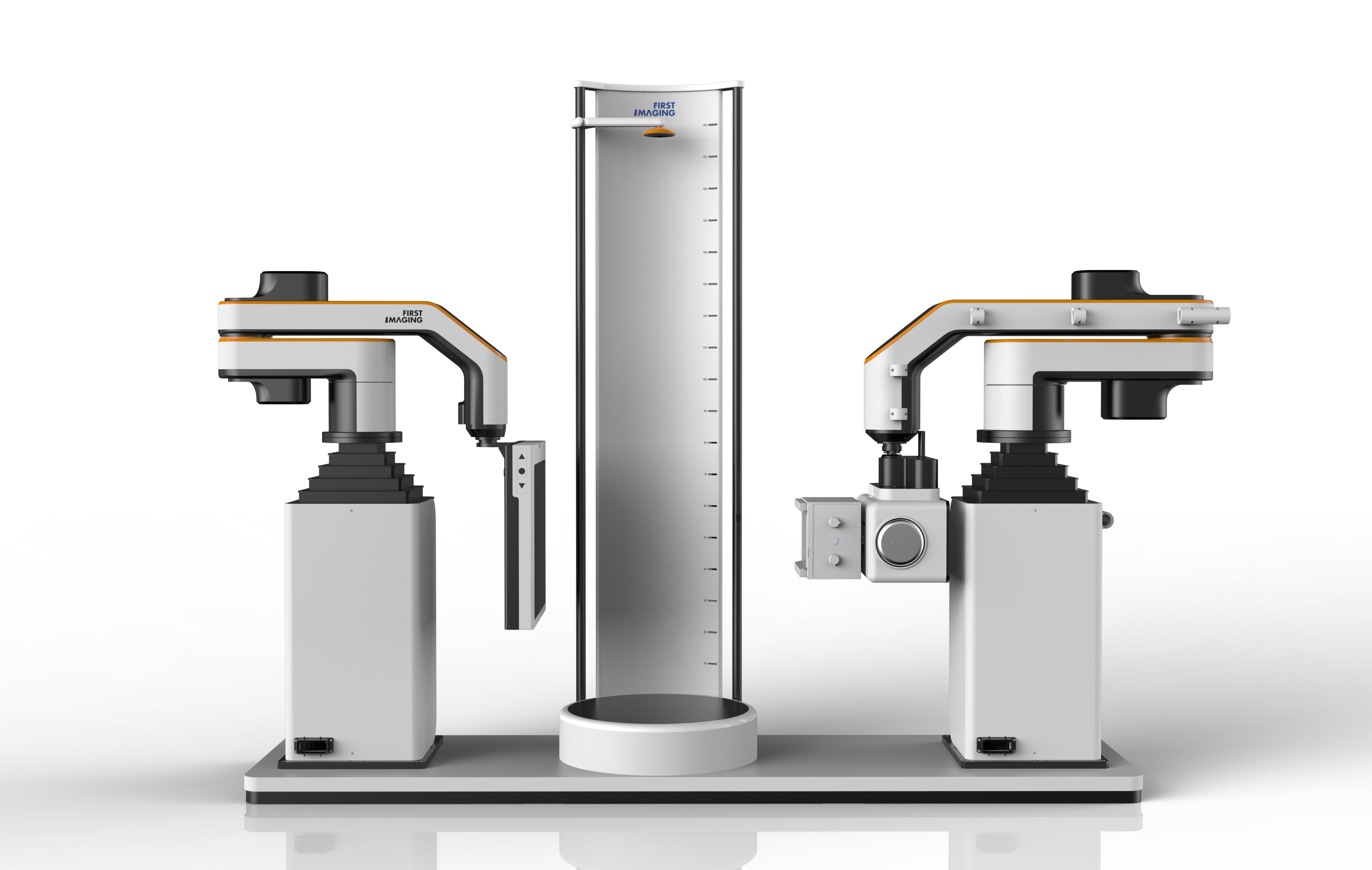}}
\caption{Robotic-arm upright position CBCT device. }
\label{fig:dev}
\end{figure}

Despite the fact that humans spend most of their day in an upright position, the most commonly used tomographic imaging techniques, including Computed Tomography (CT) and Magnetic Resonance Imaging (MRI), require the patient to be placed on a flat couch and perform imaging at a supine position. Research works show that many symptoms are more remarkable in an upright position versus a supine position \cite{jinzaki2020development}, and that diagnostic tomographic imaging at an upright position is important in assessing 3-D relationships of the bone in the ankle or spine with full weight-bearing \cite{ortolani2021angular, richter2014pedcat}, as well as measuring lung or lobe volume with gravity considered \cite{yamada2020comparison}. 

To achieve upright position imaging, one intuitive approach involves laying down conventional CT systems and having the patient standing at the central position during the scanning process. However, CT scanners usually have a limited field of view (FoV) in the z-direction. In conventional supine scanning, radiologists can manipulate the couch forward and backward for full-length-body imaging, but this is not accessible when the patient is upright. 
Contemporary systems lift the CT scanners up and down using linear motion rails and ball screws \cite{jinzaki2020development}. Due to the substantial weight of CT scanners, the stability of the floor may be challenged with these systems, potentially giving rise to safety concerns. In response to the aforementioned concerns, a robotic arm Cone Beam CT (CBCT) system was developed. 
Two robotic arms are utilized in the system to handle the X-ray tube and the flat-panel detector separately. By controlling the translation and rotation of the source and detector with these arms, arc and spiral-like scanning trajectories are made possible. This device has a reduced weight compared to conventional CT systems and is capable of performing full-body-length scans with spiral-like trajectories, offering new imaging possibilities.

Robotic-arm-based systems introduce new problems from several aspects. First, complete circular scans are unattainable with robotic arms. 
Images reconstructed from the collected data suffer from limited-angle artifacts. Second, the imaging field of view in the z-direction of the flat panel detector remains limited, necessitating the robotic arm to traverse a 180-degree arc trajectory in a reciprocating motion to encompass the entire body. For safety considerations, the source and detector cannot rotate at a high speed, while larger pitches will introduce severe artifacts due to data incompleteness. 
Third, the reconstruction of images using this geometry has never been investigated. 
In the proposed trajectory, the widely accepted FDK algorithm \cite{feldkamp1984practical} cannot be directly applied because of the inhomogeneous sampling in the z-direction. Iterative reconstruction methods\cite{trampert1990simultaneous, Sanctorum_2021, gregor2008computational} are capable of producing corresponding images under this geometry, but the reconstruction process would be extremely time-consuming due to the massive quantity of projection data. 

In this paper, we comprehensively analyzed the Full-length-body CBCT imaging model in upright position with robotic arm systems, and propose 1) a mathematical model to calculate the optimal helical pitch of the trajectory; 2) an FDK-style method with normalized backprojection for approximate and fast reconstruction. Extensive experiments demonstrate that the proposed trajectory and reconstruction algorithm are capable of generating high-quality images via experiments conducted on simulated data. The proposed methods have the potential to advance the clinical applications of upright CBCT scans.


\section{Method}
\label{sec:method}
\begin{figure}[tb]
\centerline{\includegraphics[width=0.4\textwidth]{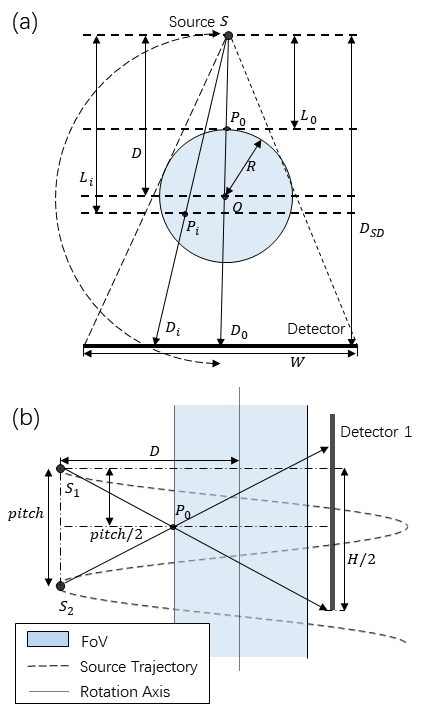}}
\caption{The proposed half-spiral scanning trajectory on (a) transversal view and (b) sagittal view. }
\label{fig:fov}
\end{figure}
\subsection{Optimal Trajectory}
\label{subsec:traj}
In the proposed scanning system, the source and the detector reciprocate on a semicircular trajectory, while descending uniformly in z-direction(see Fig. \ref{fig:fov}). 
The main issue in the trajectory design is the relationship between the rotation speed and the descent speed, which is usually determined by the helical pitch in spiral CT scanners. In our study, the pitch refers to the distance between X-ray sources before and after two entire semicircular scans(see Fig. \ref{fig:fov}(b)). The setting of the helical pitch is directly proportional to the scanning speed. However, an excessively large helical pitch can lead to insufficient data acquisition and increased artifacts. In this section, we try to find the optimal pitch value that maintains image quality while minimizing scanning time. 

In the proposed method, each voxel in the imaging FoV is supposed to be covered with scans from, at least, a source trajectory segment of 180 degrees, providing as much information as possible. To achieve this, the imaging FoV should first be determined. As most circular/helical CT scans do, the proposed system has a circular FoV on the transversal view. As depicted in Fig. \ref{fig:fov}, the radius of the circular FoV $R$ can be defined using detector width and system distances with the following equation, 
\begin{equation}
R = \frac{D\cdot W}{2\cdot \sqrt{D_{SD}^{2}+\frac{W^{2}}{4}}}, 
\end{equation}
where $D$ stands for the distance between the X-ray source and the rotation axis, $D_{SD}$ is the distance between the X-ray source and the detector and $W$ represents the width of the detector. 

Voxels closer to the X-ray source receive a larger magnification factor, making the corresponding projection points easier to get outside the detector. Referring to Fig. \ref{fig:fov} (a) as an example, $P_0$ is the point closest to the source inside the FoV while $P_i$ is a random point in the FoV. The relationship between the magnification factors $M_0$ and $M_i$ corresponding to $P_0$ and $P_i$ can be expressed with the following equation
\begin{equation}
    M_0=\frac{D_{SD}}{L_0} \geq \frac{D_{SD}}{L_i}=M_i. 
\end{equation}
It is evident from Fig. \ref{fig:fov} that the points located on the center plane of $S_1$ and $S_2$ have a greater probability of evading the detector on both views. To guarantee that all voxels are scanned from a minimum of 180 degrees, we focus on the point $P_0$, which is the most challenging point to satisfy this requirement. Geometric analysis leads to the following equation
\begin{equation}
    \frac{H}{2} \geq |z_{P_0}-z_{S_1}| \cdot M_0 = \frac{pitch \cdot D_{SD}}{2L_0} = \frac{pitch \cdot D_{SD}}{2(D-R)},
\end{equation}
which further leads to
\begin{equation}
    pitch \leq H\cdot (D-R) / D_{SD}.
\end{equation}

\subsection{Image Reconstruction}

The optimal trajectory in section \ref{subsec:traj} ensures that each voxel point inside the FoV receives at least one count at each rotation angle, but there is usually more than one. Actually, the number of counts assigned to each voxel varies depending on the rotation angle and voxel location, 
and the counts should be normalized during the reconstruction to avoid inaccuracies and artifacts in the results. The normalization is performed during backprojection with the following equation
\begin{equation}
    bp(\boldsymbol{x}, \theta)=\frac{D}{D-S}\cdot \frac{\sum_{i=1}^{K_{\theta}} F_{i,\theta}(u_{i,\theta}(\boldsymbol{x}),v_{i,\theta}(\boldsymbol{x}))}{\sum_{i=1}^{K_{\theta}} \boldsymbol{1}_{i,\theta}(u_{i,\theta}(\boldsymbol{x}),v_{i,\theta}(\boldsymbol{x}))}, 
\end{equation}
where $\boldsymbol{x}$ is a voxel, $S$ is the distance between point $\boldsymbol{x}$ and the virtual detector plane placed at the rotation axis, $K_{\theta}$ is the total number of views at rotation angle $\theta$, $F_{i,\theta}(\cdot)$ is the pre-processed projection view at the $i$-th view at rotation angle $\theta$, $u_{i,\theta}(\boldsymbol{x})$ and $v_{i,\theta}(\boldsymbol{x})$ are the piercing coordinate on the detector corresponding to point $\boldsymbol{x}$, $\boldsymbol{1}_{i,\theta}(\cdot)$ is an indicator function of the following form
\begin{equation}
    \begin{split}
        \boldsymbol{1}_{i,\theta}(u,v)= \left \{
        \begin{array}{ll}
             1, & |v| \leq H/2 \\
             0, & |v| > H/2
        \end{array}
        \right. .
    \end{split}
\end{equation}

Integrating short-scan Parker weighting and normalized backprojection into FDK reconstruction, the final reconstruction algorithm can be expressed as
\begin{equation}
    I(\boldsymbol{x})=\int_{\theta=0}^{\pi}\frac{wD}{D-S}\frac{\sum_{i=1}^K f(u(\boldsymbol{x}),v(\boldsymbol{x})) * g(u(\boldsymbol{x}))}{\sum_{i=1}^K \boldsymbol{1}(u(\boldsymbol{x}),v(\boldsymbol{x}))}d\theta, 
    \label{eq:recon}
\end{equation}
where $w=D/\sqrt{D^2+u^2+v^2}$ is the cosine weight in the FDK algorithm, $f(u,v)$ is the original projection data and $g(u)$ represents the ramp filter. For simplicity reasons, some subscripts are removed in Eq. \ref{eq:recon}. 





\section{Experiments and Results}
\label{sec:exp}
The proposed geometry and reconstruction method has been tested on simulated data. Shepp-Logan phantom, foam-like phantom from \cite{pelt2022foam} and the CT images from Visible Human Project (VHP) \cite{ackerman1995accessing} were used for data simulation. Most geometrical parameters in the simulation were set according to a prototype robotic-arm CBCT device developed and installed at First-Imaging Tech., Shanghai, China. The specific parameters are as follows: $D=412.5mm$, $D_{SD}=1100mm$ and $W=H=430mm$. The FoV radius $R=79.13mm$ and optimal pitch $pitch=130.32mm$ are computed according to section \ref{subsec:traj}. 




\subsection{Experiments on trajectory}
\begin{figure}[t]
\centerline{\includegraphics[width=0.5\textwidth]{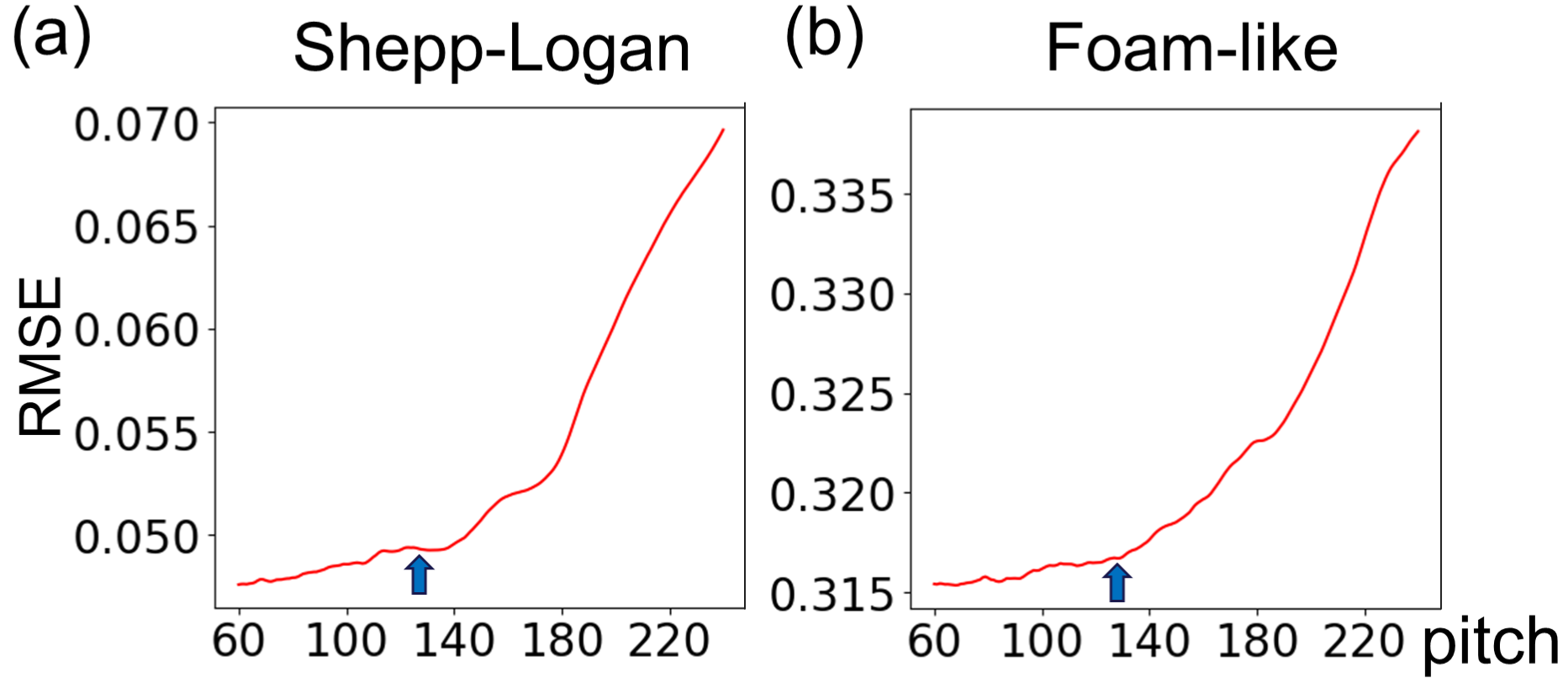}}
\caption{RMSE curve with respect to pitch settings on (a) Shepp-Logan phantom and (b) Foam-like phantom. }
\label{fig:rmse}
\end{figure}

To validate the optimal pitch, experiments were conducted with various pitch settings on the Shepp-Logan phantom and the foam-like phantom. Root Mean Squared Error (RMSE) values between the original volumes and the reconstructed volumes from the proposed method were calculated to evaluate imaging accuracy. The RMSE curves with respect to different pitch settings are depicted in Fig. \ref{fig:rmse} and the optimal pitch is marked out with blue arrows. 

In general, as the pitch increases, so does RMSE. The RMSE values exhibit a gradual increase prior to attaining the calculated optimal pitch, which can be attributed to the cone-beam artifact. Once the optimal pitch is surpassed, the RMSE increases at an accelerated rate due to the severe image degradation caused by inadequate data collection. 

\subsection{Comparison with other reconstruction methods}
\begin{figure}[tb]

\centerline{\includegraphics[width=0.45\textwidth]{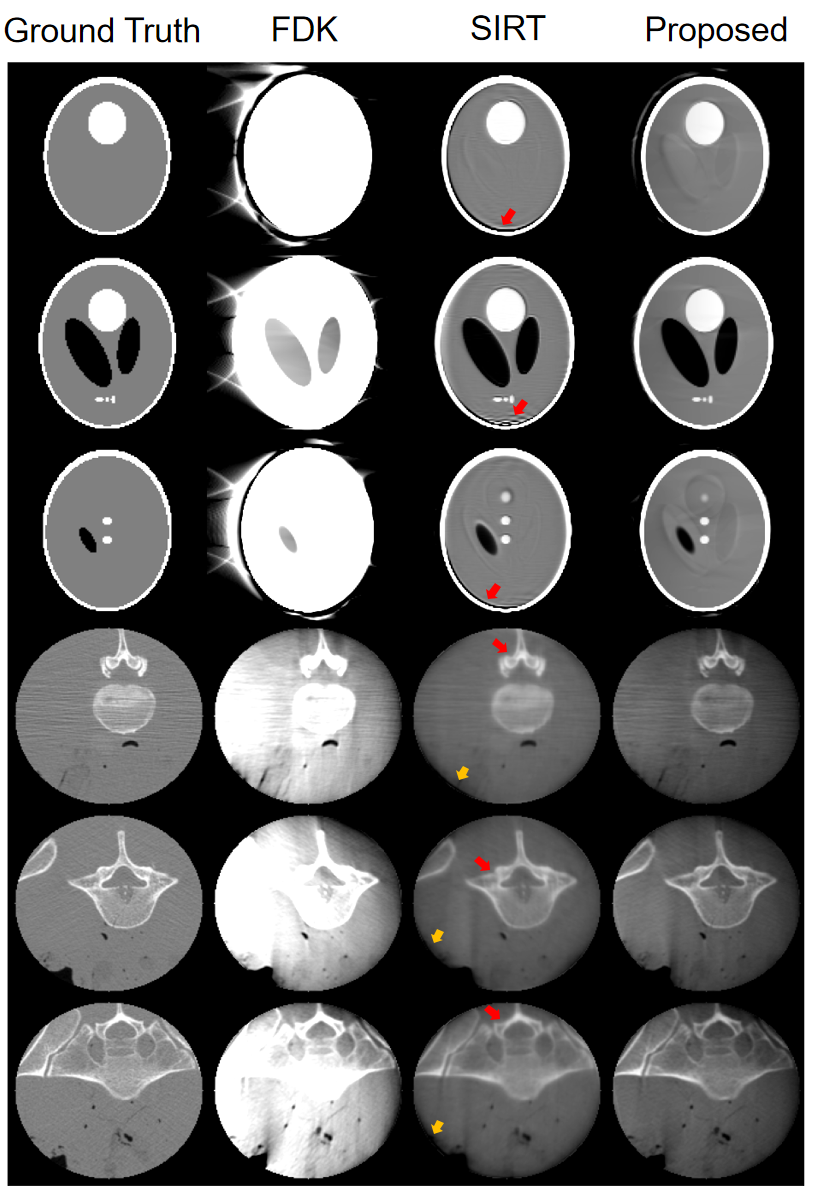}}
\caption{Experimental results on testing data.}
\label{fig:shlo}
\end{figure}


The proposed reconstruction method is compared with the widely accepted FDK algorithm and the Simultaneous Iterative Reconstruction Technique (SIRT) on all phantoms. The FDK algorithm used here comprises the short-scan Parker weighting to reduce intensity drop but not the normalized backprojection. The SIRT algorithm undertook 200 iterations and volumes with a minimal RMSE value were selected as the final results (usually achieved after 130-150 iterations). 

Fig. \ref{fig:shlo} gives visual comparisons between different methods on Shepp-Logan phantom and VHP data. Due to normalization issues, the results from the FDK algorithm suffer from substantial value shifts, making them unacceptable for clinical use. SIRT yields much more consistent values, but tissue boundaries in the images are often over-smoothed (see red arrows in Fig. \ref{fig:shlo}). Furthermore, artifacts arising from data truncation become more severe as the iteration proceeds (see orange arrows in Fig. \ref{fig:shlo}). The proposed method is able to reconstruct images with accurate values and the boundaries are much sharper than those from SIRT. 

Quantitative results are presented in Table \ref{tab:quan}. The proposed method achieves the best RMSE values on the Shepp-Logan phantom, while SIRT outperforms ours on the foam-like phantom and VHP data. As for computation time, SIRT takes over 11 minutes on foam data and 31+ minutes on VHP data. The proposed method uses 1-3 minutes for reconstruction and can be further accelerated through source code optimization. Notice that all experiments were performed on a laptop with Intel(R) Core(TM) i9-13900HX CPU and NVIDIA GeForce RTX 4060 Laptop GPU. The algorithms were implemented in MATLAB with forward/backward-projection for each view accelerated with CUDA.



\begin{table}[tb]
    \caption{Quantitative comparison between different reconstruction algorithms}
    \centering
    \resizebox{0.95\columnwidth}{!}{
    \begin{tabular}{cc|cccc}
      \toprule[0.5mm]
      &Phantom & Shepp-Logan & foam-like & VHP \\
      \midrule
      \midrule
      \multirow{4}{*}{\rotatebox{90}{\textit{RMSE}}}
      &FDK & 0.1691  & 0.5588 & 319.0632 \\
      &SIRT & 0.0713  & \textbf{0.2348} & \textbf{161.0095} \\
      \cmidrule{2-5}
      & Proposed &\textbf{0.0492} & 0.3168 & 211.2213\\
      \midrule
      \midrule
      \multirow{4}{*}{\rotatebox{90}{\textit{time(S)}}}
      &FDK & \textbf{2.4573 }  & \textbf{4.9335 } & \textbf{9.9501} \\
      &SIRT & 272.3& 679.2034 & 1903.4910 \\
      \cmidrule{2-5}
      & Proposed & 45.3434 & 63.9502  & 189.3648\\
      \bottomrule[0.5mm]
    \end{tabular}
    }
    \label{tab:quan}
\end{table}

\section{Discussion and Conclusion}
\label{sec:diss}
Tomographic imaging at an upright position is of great significance for functional analysis of orthopedics, especially for postoperative analysis of spine and lower limb orthopedics. Robotic-arm-based upright position CBCT systems have potential to become the optimal choice for this task for several reasons: 1) CBCT usually has a lower radiation dose to the human body than CT; 2) robotic arm systems have lower environmental requirements and are safer to patients than the upright CT/MRI systems; 3) orthopedic imaging does not require high soft tissue contrast, which just makes up for the weaknesses of CBCT imaging.

Several simulated data experiments prove the efficacy of the proposed method. Despite that the geometrical parameters utilized in our experiments were obtained from a genuine system, we did not perform experiments on real-scanned data because the phantom for precise geometrical calibration is still under production. Due to the system's objective of long-range imaging in the z-direction, traditional short phantoms are incapable of calibration along the entire trajectory. 

The proposed method cannot alleviate cone beam artifacts and limited-angle artifacts, which is the main reason why the proposed method RMSE is inferior to the iterative algorithm. However, iterative algorithms have their own shortcomings. Since each iteration requires a forward and back projection, the algorithm will be very time-consuming. Furthermore, the mismatch between forward and back projection will introduce additional artifacts. The proposed method can also be used to initialize the iterative algorithm, reducing the number of iterations before convergence.

To conclude, this paper dives into the system design and image reconstruction for robotic-arm CBCT in the upright position, and the proposed methods have the potential to contribute to the current state of research in upright position imaging and functional analysis of orthopedics.

\section{Acknowledgement}
This work is supported in part by First-Imaging Technology Co. Ltd., Shanghai, China. The algorithms developed have been integrated into the \textit{first3D} orthopedic imaging software and work along with the Huashan Upright Position Digital Imaging System by First-Imaging. This work is also supported in part by the State Key Project of Research and Development Plan under Grants 2022YFC2401600 and 2022YFC2408500, in part by the National Natural Science Foundation of China under Grant T2225025, and in part by the Key Research and Development Program of Zhejiang Province under Grant 2021C03029. 

\section{Compliance with Ethical Standards}
This is a numerical simulation study for which no ethical approval was required.

\bibliographystyle{IEEEbib}
\bibliography{refs}

\begin{thebibliography}{10}

\bibitem{jinzaki2020development}
Masahiro Jinzaki, Yoshitake Yamada, Takeo Nagura, et~al.,
\newblock ``Development of upright computed tomography with area detector for whole-body scans: phantom study, efficacy on workflow, effect of gravity on human body, and potential clinical impact,''
\newblock {\em Investigative radiology}, vol. 55, no. 2, pp. 73, 2020.

\bibitem{ortolani2021angular}
Maurizio Ortolani, Alberto Leardini, Chiara Pavani, et~al.,
\newblock ``Angular and linear measurements of adult flexible flatfoot via weight-bearing ct scans and 3d bone reconstruction tools,''
\newblock {\em Scientific Reports}, vol. 11, no. 1, pp. 16139, 2021.

\bibitem{richter2014pedcat}
Martinus Richter, Bernd Seidl, Stefan Zech, and Sarah Hahn,
\newblock ``Pedcat for 3d-imaging in standing position allows for more accurate bone position (angle) measurement than radiographs or ct,''
\newblock {\em Foot and Ankle surgery}, vol. 20, no. 3, pp. 201--207, 2014.

\bibitem{yamada2020comparison}
Yoshitake Yamada, Minoru Yamada, Shotaro Chubachi, et~al.,
\newblock ``Comparison of inspiratory and expiratory lung and lobe volumes among supine, standing, and sitting positions using conventional and upright ct,''
\newblock {\em Scientific reports}, vol. 10, no. 1, pp. 16203, 2020.

\bibitem{feldkamp1984practical}
Lee~A Feldkamp, Lloyd~C Davis, and James~W Kress,
\newblock ``Practical cone-beam algorithm,''
\newblock {\em Josa a}, vol. 1, no. 6, pp. 612--619, 1984.

\bibitem{trampert1990simultaneous}
Jeannot Trampert and Jean-Jacques Leveque,
\newblock ``Simultaneous iterative reconstruction technique: Physical interpretation based on the generalized least squares solution,''
\newblock {\em Journal of Geophysical Research: Solid Earth}, vol. 95, no. B8, pp. 12553--12559, 1990.

\bibitem{Sanctorum_2021}
Joaquim~G Sanctorum, Sam~Van Wassenbergh, Van Nguyen, et~al.,
\newblock ``Extended imaging volume in cone-beam x-ray tomography using the weighted simultaneous iterative reconstruction technique,''
\newblock {\em Physics in Medicine \& Biology}, vol. 66, no. 16, pp. 165008, 2021.

\bibitem{gregor2008computational}
Jens Gregor and Thomas Benson,
\newblock ``Computational analysis and improvement of sirt,''
\newblock {\em IEEE Transactions on medical imaging}, vol. 27, no. 7, pp. 918--924, 2008.

\bibitem{pelt2022foam}
Dani{\"e}l~M Pelt, Allard~A Hendriksen, and Kees~Joost Batenburg,
\newblock ``Foam-like phantoms for comparing tomography algorithms,''
\newblock {\em Journal of Synchrotron Radiation}, vol. 29, no. 1, pp. 254--265, 2022.

\bibitem{ackerman1995accessing}
Michael~J Ackerman,
\newblock ``Accessing the visible human project,''
\newblock {\em D-Lib Magazine}, vol. 1, no. 4, 1995.

\end{thebibliography}

\end{document}